\documentstyle[prb,aps,twocolumn,epsfig]{revtex}
\title{Magnetomechanics of mesoscopic wires}  
 
\author{Sara Blom}
\address{Department of Applied Physics, Chalmers University of
Technology and G\"oteborg University, SE-412 96 G\"oteborg, Sweden}

\begin{document}
\narrowtext
\maketitle
 
\begin{abstract}
We have studied the force in mesoscopic wires in the presence of an external
magnetic field along the wire using a free electron model.  We show that
the applied magnetic field can be used to affect the force in the wire. 
The magnetic field breaks the degeneracy of  the eigenenergies of the conduction modes, resulting in more structure in the force as a function of wire length. The use of an external magnetic field is an equilibrium method to control
the number of transporting channels.
Under the least favorable circumstances (on the middle of a low
conduction step) one needs about 1.3 T, for a mesoscopic Bismuth wire, to see an
abrupt change in the force, at fixed wire length.
\end{abstract}
 
\section{Introduction}
The electrical conductance in a ballistic wire with dimensions comparable to the
Fermi wavelength increases in steps of $G_0=2e^2/h$ as the cross section
increases. This conductance quantization is observable at room
temperature in metallic nanowires formed by pressing two pieces of metal
together, into a metallic contact. When the two pieces are separated the
contact is stretched into a nanowire, a wire of nanometer dimensions.
Several experiments varying this principle have been performed, e.g.
using scanning tunneling microscopy\cite{Pascual93}, mechanically
controlled break junctions\cite{Muller92} or just plain macroscopic
wires~\cite{Costa-K95}. Most nanowire experiments have been performed
on metals, however conductance quantization have been seen~\cite{Costa-K97} in Bismuth at
4K. Since Bismuth has a Fermi wavelength $\lambda_F=26$
nm~\cite{Costa-K97}, these semi-metal ``nanowires'' are larger than the
metallic nanowires.
 
The stepwise variation of conductance in such a mesoscopic wire is
accompanied by an abrupt change of the force in the wire~\cite{Rubio96}.
Using a free electron model, neglecting all atomic structure of the
wire, it has been
shown\cite{Stafford97,Ruitenbeek97,Yannouleas97,Blom98} that the size of
the electronic contribution to the force fluctuations are comparable to
the experimentally found values and that the qualitative behavior, i.e.
the abrupt change that accompanies the conductance steps, is the same.
 
In the wire the transverse motion of the electrons give rise to
quantized modes $\alpha$ of energy $E_\alpha$. In the simplest version of the
Landauer formalism, a mode is considered fully transmitting, open, if
$E_F>E_\alpha$ and closed otherwise.\cite{Dittrich} Each open mode contributes an amount
$e^2/h$ to the conductance, if modes with different spin are considered separately. When the wire is
elongated, the cross section decreases,  more and more modes are pushed
above the Fermi level and closed, thus decreasing the conductance
stepwise. This has been shown in two dimensions\cite{Glazman} and in
three dimensions\cite{Bogachek}. 
 
It has been suggested\cite{Zagoskin98} that the conductance and the mechanical force in a
nanowire can be controlled by an applied driving voltage. This effect
originates from the injection of additional electrons with voltage
dependent energy, because of the different chemical potentials of the two
reservoirs. A relatively large applied voltage is needed so one will
have to worry about heating in this case. 
 
The eigenenergies of the transverse motion can be affected by an
external magnetic field, $B$, perpendicular to the cross section of the
wire. This will show in the conductance and in the force as a function
of $B$. The effect of a magnetic field on the conductance has been considered in ref. \onlinecite{Bogachek2}. To use an external magnetic field is an equilibrium method to control
the number of transporting channels, without significant risk of
relaxation. 
 
Because of band bending, due to the small size of the wire, the eigenenergies will have to be corrected. This can, however, be taken care of by
introducing an effective Fermi energy in the wire, $\tilde{E}_F$.
Assuming that the number of electrons (per unit volume) is constant,
$\tilde{E}_F$ can be determined selfconsistently and will vary with
wire length and magnetic field. 
 
In this paper we present force calculations for different applied magnetic
fields and wire lengths, using a free electron model. We take into account
the effect 
of band bending, adjusting the Fermi energy in the wire. In order
to resolve any effect for moderate magnetic fields, a low
cyclotron effective mass (which enters in the cyclotron frequency)
is needed, which can be found in semi-metals. Metals are less favorable since because of a larger cyclotron effective mass (larger Fermi energy) we would need a larger magnetic field in order to resolve any effect. For numerical
estimates we have used values for Bismuth, a typical semi-metal. For Bismuth also the spin splitting is
important since it has a large spectroscopic spin splitting factor $g$.
 
\section{Model}
We consider a cylindrical ballistic wire of length $L$ with circular cross section and a parabolic confining potential, 
\begin{equation}
\omega(r)=\frac{\omega_0^2m^*r^2}{2}\equiv E_F\frac{r^2}{R^2},\label{eg:ldef}
\end{equation}
using cylindrical coordinates $(r,\phi,z)$ and where $m^*$ is the effective
electron mass. The wire is along the $z$-direction. The last equality in Eq.~\ref{eg:ldef} defines $\omega_0$. In this equation $E_F$ is the zero $B$-field bulk value, yielding a magnetic field independent confining potential. We assume that the volume $V=\pi R^2L$ of the wire is kept constant
during elongation, which makes $R$ and $L$  mutually dependent.
 
With the above confining potential and an applied magnetic field along
the wire the Schr\"odinger equation has been solved~\cite{Liu}. If 
also spin is included the eigenenergies are 
\begin{eqnarray}
E_\alpha & = &
\hbar\left(\frac{\omega_c^2}{4}+\omega_0^2\right)^{1/2}n+\frac{1}{2}l\hbar\omega_c+sg\mu_BB\label{eq:eig}\\
n & = & 2m+|l|+1\nonumber\\
m & = & 0,1,2,\ldots\nonumber\\
l & = & 0,\pm 1,\pm 2,\ldots\nonumber\\
s & = & \pm 1/2\nonumber\\
\alpha & = & \{m,l,s\},\nonumber
\end{eqnarray}
where $\omega_c=eB/m^*$ is the cyclotron frequency, $\mu_B$ is the
Bohr magneton and $sg\mu_B$ is the magnetic moment associated with the electronic spin.
 
Since our system is open the electronic contribution to the force in the
wire is given by the derivative of the grand potential $\Omega=E-\mu N$
with respect to elongation. Here $E$ is the total energy of the
electrons in the wire, $\mu$ the chemical potential and $N$ the number
of electrons in the wire. If the Fermi energy $E_F$ is much higher than
the thermal energy (as in metals or at low temperature) we have
$\mu\approx E_F$. The grand potential is then~\cite{Blom98}
\begin{equation}
\Omega(E_F)=-\sum_\alpha\frac{4}{3}L\sqrt{\frac{2m^*}{\pi^2\hbar^2}}(E_F-E_\alpha)^{3/2},
\end{equation}
where the sum is over all open modes. The force in the wire is given by 
\begin{equation}
F=-\frac{\delta\Omega}{\delta L},
\end{equation}
which in general has to be calculated numerically.
 
The magnetic field affects the system primarily by splitting the
otherwise degenerate eigenenergies of the conduction modes,
Eq.~\ref{eq:eig}. Since then the conduction modes will open
one by one this will cause more structure in the force and conductance
when displayed as a function of wire length. Subsequently, when applying an external
magnetic field we will see the (clearest) effect when the highest open
level or the lowest closed level goes through the Fermi level (whichever
happens first). If we do not adjust the Fermi energy for band bending,
but use the bulk Fermi energy for zero magnetic field, one can
analytically calculate the $B$-field needed, when keeping the wire at a
specific length. The least favorable situation would be on the middle of
a conduction step.
 
\section{Results and discussion}
 
We have used numerical values for Bismuth, a typical semi-metal with $E_F=25$
meV\cite{Costa-K97}. Bismuth has an anisotropic Fermi surface resulting
in different effective masses in different directions, between
$0.009m_e-1.8m_e$\cite{Galt}. The cyclotron effective mass is in the
range $0.009m_e-0.13m_e$\cite{Galt}. Assuming an isotropic Fermi 
surface and an quadratic dispersion relation, both effective masses 
are the same, for $E_F=25$ meV $m^*=0.07m_e$. The spectroscopic splitting 
factor, $g$, can be as high as 260, or one order of magnitude smaller depending on
the direction of the magnetic field\cite{Cohen}. With $g=20$ the spin
splitting is roughly of the same order as the Landau level distance, and becomes
dominant for $g$ as large as 200. We have used $g=20$. The wire 
volume was kept constant at $30000$ nm$^3$.

To find the effective Fermi energy of the wire we have adjusted the value in order 
to keep the number of electrons constant, with a tolerance of $10^{-4}\%$.
 
\begin{figure}
\epsfig{file=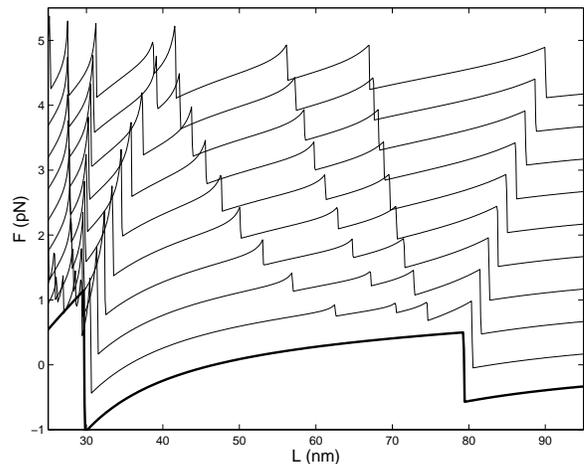, width=7.7cm}
\caption{The force in a  mesoscopic wire as a function of wire length for
different magnetic fields. The lowest, thick, curve is for $B=0$. The
next lines, each displaced by  0.5 pN, are for $B=0.5$ T, $B=1$ T etc, the uppermost line being for $B=4.5$ T.  The splitting of the eigenenergies of
the conduction modes is clearly visible: for larger $B$-fields the curves
have more structure since now every mode closes one by one when the wire is
elongated. We have used the spectroscopic splitting factor $g=20$ and an 
effective Fermi energy $\tilde{E}_F$.\label{fig:force_c}} 
\end{figure}
 
Figure \ref{fig:force_c} shows the force in the wire as a function of 
wire length for different magnetic fields. For non-zero fields the force 
curves show more structure since now the eigenenergies of the conduction 
channels are non-degenerate and close one by one, each time resulting in a 
sharp change of the force. 
 
\begin{figure}
\epsfig{file=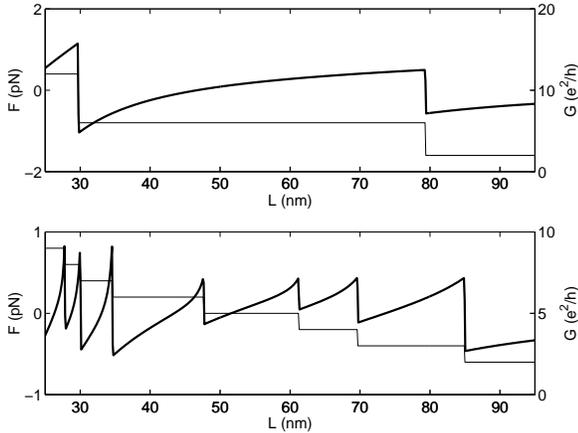, width=7.7cm}
\caption{The force (thick line) and the conductance in a mesoscopic wire for two different
magnetic fields, in the upper figure $B=0$ and in the lower figure $B=2.5$
T. We clearly see that the abrupt change in the force happens when a
channel closes, i.e. when there is a step in the conductance. We have used an 
effective Fermi energy $\tilde{E}_F$.\label{fig:foco_c}
}
\end{figure}

\begin{figure}
\epsfig{file=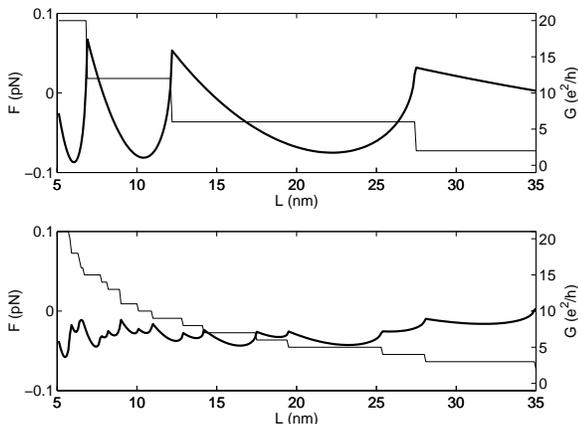, width=7.7cm}
\caption{The force (thick line) and the conductance in a mesoscopic wire for the less realistic case of a constant Fermi energy in the wire equal to the  zero $B$-field bulk value (25 meV). Results for two different magnetic fields are shown, in the upper figure $B=0$ and in the lower figure $B=2.5$ T. \label{fig:foco_a}} 
\end{figure}

The force and conductance for two particular 
magnetic fields, $B=0$ and $B=2.5$ T, are shown in Fig.~\ref{fig:foco_c}. Each 
step in the conductance is accompanied by an abrupt change in the force. We also show the corresponding picture for the simplest possible case\cite{Blom98}, when we use the bulk value of the Fermi energy, $E_F$, in Fig.~\ref{fig:foco_a}. In this case the force is one order of magnitude smaller then in the more realistic case with $\tilde{E}_F$. This is because 
the effective Fermi energy has to be larger then the bulk value in order to keep the number 
of electrons per unit volume in the wire constant in spite of the quantization 
of levels. Also the conduction modes close much later in the $\tilde{E}_F$-case than in the more simple case when the wire is elongated. The reason for this is that the effective Fermi energy, as a function of wire length, follows the eigenenergies before intercepting it and closing the channel.
 
On the middle of the second conduction step ($G=3G_0$, $n=2$)  
the circumstances 
are least favorable to see the effect of the magnetic field.
For the case with the zero $B$-field 
bulk value of the Fermi energy ($L=19.8$nm)
we have analytically calculated  that one needs $B=2.4$ T, to see the
highest open level go through the Fermi energy, thus giving a sharp
change in the force as well as in the conductance. For higher conduction 
modes one will see the effect for smaller fields, since the splitting is 
proportional to $l$, whos
absolute maximum is equal to $n$. 
 
\begin{figure}
\epsfig{file=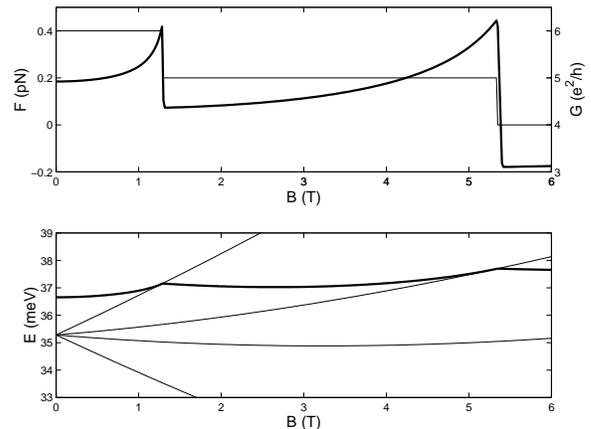, width=7.7cm}
\caption{In the upper figure we show the force (thick line) and conductance for $L=54.6$ nm. 
This length corresponds to the middle of the second conduction step. In the 
lower figure we show the eigenenergies of the second conduction step and the 
effective Fermi energy of the wire (thick line). We see that when the highest 
level goes through the Fermi level (for approximately $B=1.3$ T) there is a step 
in the conductance and an abrupt change in the force. \label{fig:fix_c}} 
\end{figure}

In Fig.~\ref{fig:fix_c} we see the 
force and the conductance as a function of magnetic field for a fixed wire 
length, $L=54.6$ nm. This is for the case with an effective wire Fermi energy and 
the length corresponds to  the middle of the second conduction step 
($G=3G_0$, $n=2$). We see that we need about 1.3 T before the highest open 
level goes through the Fermi surface showing us the articulate effect of 
the magnetic field. In the lower part of the same figure we also see the effective Fermi energy 
(thick line) and the eigenenergies of the second conduction steps. Notice how 
the Fermi level increases with the eigenenergy before it intercepts. These 
variations are however small compared to the overall magnitude of the Fermi energy.

\begin{figure}
\epsfig{file=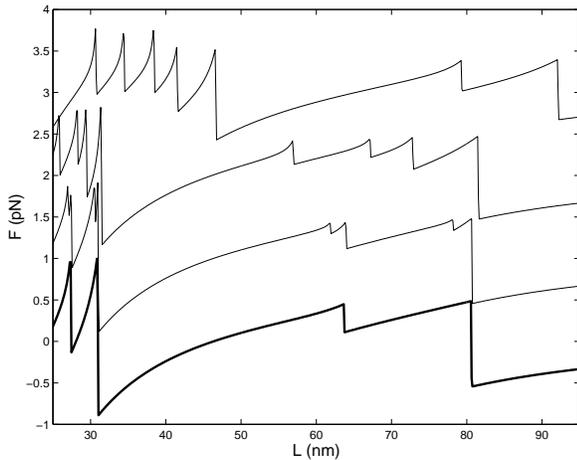, width=7.7cm}
\caption{Force as a function of length for $B=1$ T for different $g$-factors. The lowest curve is for $g=0$, and the following curves, each displaced by 1 pN, for $g=2$, $g=20$ and $g=200$ respectively. We see that no matter what the $g$-factor is, an external magnetic field will give the force curves more structure than for $B=0$, cf. Fig.~\ref{fig:force_c}.\label{fig:lande}}
\end{figure}

So far we have used the spectroscopic splitting factor $g=20$. In Fig.~\ref{fig:lande} 
we show the force as a function of length for $B=1$ T for different $g$-factors: 
$g=0,2,20$ and 200. For $g=0$ there is no spin splitting, but we still see more structure than 
for $B=0$ (cf. Fig.~\ref{fig:force_c}). This is due to the breaking of the degeneracy into the Landau levels. With 
increasing $g$-factor the spin splitting becomes larger and larger, however whatever the size of the 
spin-splitting is: more structure in the force appears with an applied magnetic field.

Also the Fermi energy of the bulk will be affected by the magnetic
field, due to the de Haas-van Alphen effect. In the case when an effective Fermi energy, $\tilde{E_F}$, is used this does not affect the results since the bulk Fermi energy does not enter into the calculations. When adjusting the bulk Fermi energy for de Haas-van Alphen effect, in the more simple case shown in Fig.~\ref{fig:foco_a}, there is no significant change on the force. We have also studied the influence of a moderate applied voltage (in the mV-range) but have seen no significant effect.
 
For metals the Fermi energy is in the eV-range demanding a much higher
magnetic fields to resolve results similar to those for Bismuth above. Since
the size of the splitting is proportional to the number of open
channels, having more channels will decrease the magnetic field needed. So if we design the circumstances to be more favorable, i.e. more open channels and close to a conduction step a moderate magnetic field will be enough to make an eigenenergy go through the Fermi level, thus giving an effect in the force and in the conductance.

\section{Conclusion}
Using a free electron model we have shown that the force in a mesoscopic wire
can be affected by an external magnetic field parallel to the wire. 
With a magnetic field present the degenerate eigenenergies of
the conduction modes split and become conducting, open, at different elongations resulting
in more force fluctuations with increasing wire length. At fixed wire
length we propose that an external magnetic field is an equilibrium method that can be used to affect
the force as well as the conductance in mesoscopic wires.

I wish to thank Robert Shekhter for valuable discussions. Financial support from the Swedish NFR is gratefully acknowledged.

\end{document}